\documentclass{svjour3}                     % onecolumn (standard format)
\smartqed  % flush right qed marks, e.g. at end of proof
\usepackage{graphicx}
\usepackage{fix-cm}
\usepackage{mathptmx}      % use Times fonts if available on your TeX system
\usepackage{hyperref}
\hypersetup{
pdfborder={0 0 0}, %omits the borders for links
colorlinks=true, % false: boxed links; true: colored links
citecolor=blue,
urlcolor=blue
}
\journalname{}
\begin{document}

\title{Dynamical scaling of the critical velocity for the onset of turbulence in oscillatory superflows
}

\titlerunning{Dynamical scaling of the critical velocity in oscillatory superflows}        % if too long for running head

\author{R. H\"anninen         \and
        W. Schoepe %etc.
}


\institute{R. H\"anninen \at
              Low Temperature Laboratory, Aalto University, P.O. Box 15100, FI-00076 AALTO, Finland \\
              \email{Risto.Hanninen@tkk.fi}           %  \\
           \and
           W. Schoepe \at
              Fakult\"at f\"ur Physik, Universit\"at Regensburg, D-93040 Regensburg, Germany
}

\date{March 25, 2011}

\maketitle

\begin{abstract}

The critical velocity $v_c$ for the onset of turbulence in oscillatory flows of superfluid helium is known to depend on the oscillation frequency $\omega$, namely $v_c\sim\sqrt{\kappa\omega}$ where $\kappa$ is the circulation quantum. Only the numerical prefactor may have some geometry dependence. This universal behaviour was described earlier qualitatively either by employing the superfluid Reynolds number or by extending known dc vortex dynamics to ac flow.  In our present work we emphasize that $v_c(\omega)\propto \sqrt{\omega}$ can also be derived rigorously by means of dynamical scaling of equations of vortex dynamics as pointed out by Kotsubo and Swift already two decades ago. 

\end{abstract}

\section{Introduction}

Dynamical scaling of an equation of motion is a simple but powerful tool for deriving relations between various parameters of that system. For example, in a classical viscous liquid the Navier-Stokes equation

\begin{equation}
\mathbf{\partial}_t\mathbf{v} + ( \mathbf{v}\cdot\nabla)\mathbf{v} = -\frac{1}{\rho}\nabla p + \nu\nabla^2\mathbf{v}
\label{e.critvel}
\end{equation}
is invariant when lengths are scaled by $l = \lambda \, l^{\ast}$, velocities by $v = v^{\ast}/\lambda$, and times by $t = \lambda^2 \,t^{\ast}$ \cite{Frisch}. This scaling also leaves the Reynolds number unchanged
\begin{equation}
Re = \frac{lv}{\nu} = \frac{l^{\ast}v^{\ast}}{\nu},
\end{equation}
which is the famous similarity principle of fluid dynamics.\\ In the following we will summarize what is known about dynamical scaling of the motion of vortex lines in superfluid helium. 

\section{Scaling of vortex dynamics}
In his seminal paper of 1988 Schwarz \cite{Schwarz} has scaled his equation of motion of a vortex filament $\mathbf{s} = \mathbf{s}(\xi,t)$\\
\begin{eqnarray}
\mathbf{\dot{s}}
 = \beta \mathbf{s^{\prime}} \times \mathbf{s^{\prime\prime}} + \mathbf{v}_s + \alpha \mathbf{s^{\prime}} \times (\mathbf{v}_{ns} - \beta \mathbf{s^{\prime}}\times \mathbf{s^{\prime\prime}}) \nonumber\\
- \alpha^{\prime}\mathbf{s^{\prime}}\times [\mathbf{s^{\prime}}\times (\mathbf{v}_{ns} - \beta \mathbf{s^{\prime}}\times \mathbf{s^{\prime \prime}})],
\label{e.vL}
\end{eqnarray}
where the primes denote derivatives with respect to the arc length $\xi$, $\alpha$ and $\alpha'$ are the parameters of mutual friction, and $ \beta = (\kappa/4\pi) \ln{(cR/a_0)} $, with $c \sim$ 1, $R$ is an effective radius of curvature of the filament, and $a_0$ the core radius. The counterflow velocity, $\mathbf{v}_{ns} = \mathbf{v}_n - \mathbf{v}_s$, is simply the difference between normal and superfluid velocities. After introducing reduced time and velocity scales $t_0 = \beta \, t$ and $v_0 = v/\beta$ the resulting equation was shown to be invariant under the same scaling as discussed above.

Even if the above scaling was derived by using a so called local induction approximation (LIA) the same applies also when using a full Biot-Savart equation (BSE). The inclusion of the non-local term, which itself satisfies the velocity scaling, only slightly modifies the local term. Since, as will be noted below, the corrections due to the logarithmic term are small and very difficult to observe in experiments, we ignore here the small changes in the logarithmic term that are due to differences between LIA and BSE.

It should be mentioned that from Eq.(\ref{e.vL}) Schwarz calculated a rate equation of the change of the vortex density $dL/dt_0$ due to a growth term $L^{3/2}$ and a loss term $L^2$ that is scale invariant and that corresponds to the well known Vinen equation \cite{VinenEq}, see below.

\section{Scaling of oscillating superflows}
\begin{figure}[ht]
\centerline{\includegraphics[width=0.75\columnwidth,clip=true]{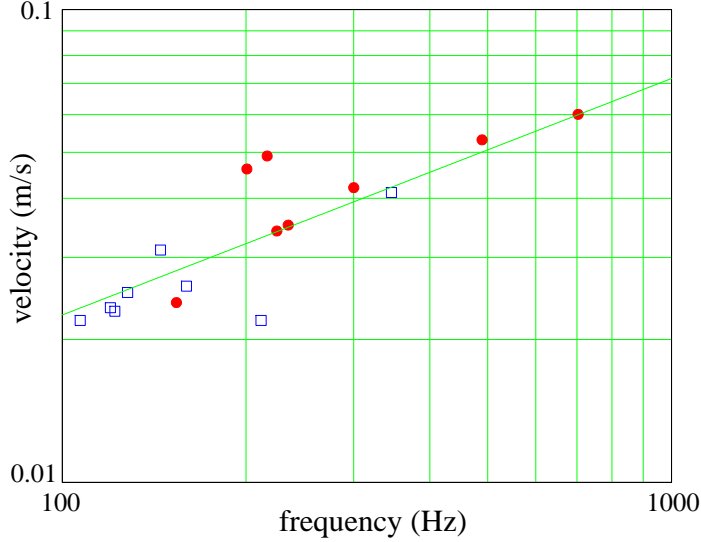}}
\caption{\label{f:vcsphere}(From Ref.\cite{arxiv}) Critical velocity for the onset of turbulence as a function of the frequency of 2 oscillating spheres. Blue squares: radius $124 \mu$m; red dots: radius $100 \mu$m; green solid line: $v_c=2.85\,\sqrt{\kappa 2\pi\!f}$ (m/s); temperature is below 1 K.}
\end{figure}

In ac flow experiments the oscillation frequency $\omega$ is a scaling variable. This has been recognized some time ago by Kotsubo and Swift when analyzing their experiments on vortex turbulence generated by high-amplitude second sound in helium-4 \cite{Swift1,Swift2}. In the following, we apply their scaling results to our experiments on ac flows generated by an oscillating sphere. From the work of Schwarz \cite{Schwarz} they find that the time scale $1/\omega$ transforms as 

\begin{equation}
\omega_0 = \lambda^{-2} \omega_0^{\ast},
\label{e.omega}
\end{equation}
where $\omega_0 = \omega/\beta$ and $\omega_0^{\ast} =\omega^{\ast}/\beta^{\ast}$ with $\beta^{\ast} = (\kappa/4\pi) \ln{(cR^{\ast}/a_0)}$. From $v_{c0} = \lambda ^{-1} v_{c0}^{\ast}$ they obtain by eliminating $\lambda$ as a final result

\begin{equation}
v_c/v_c^{\ast} = (\beta/\beta^{\ast})^{1/2} (\omega/\omega^{\ast})^{1/2}.
\label{e.vc}
\end{equation}
This is how the square-root dependence of $v_c(\omega)$ is derived. There is, however, the logarithmic correction by $(\beta/\beta ^{\ast})^{1/2}$ due to the scaling $R = \lambda R^{\ast}$. We can estimate the logarithmic corrections from the definition of $\beta$ and $\beta ^{\ast}$ and using Eq.(\ref{e.omega}) by solving for $\beta/\beta ^{\ast}$ \cite{Swift2}
\begin{equation}
\beta/\beta^{\ast} = 1 + \frac{\ln[(\beta/\beta^{\ast})^{1/2}(\omega ^{\ast}/\omega)^{1/2}]}{\ln{(cR^*/a_0)}}.
\end{equation}
If we assume that $R$ and $R^{\ast}$ are approximately given by the intervortex spacing as suggested by Schwarz \cite{Schwarz2} we estimate as an order of magnitude $\ln{(cR^{\ast}/a_0)} \approx$ 10. Inserting this we find $(\beta/\beta^{\ast})^{1/2}$ = 1.04 for our complete frequency range $\omega^{\ast}/\omega \approx$ 7. This is a small correction that could not be resolved in our experiment, see Fig.\ref{f:vcsphere} \cite{arxiv}.

It is true that in our experiments the sphere radius is kept fixed and therefore the square-root dependence of $v_c$ is not an automatic result from scaling. But as noted by Kotsubo and Swift, the scaling suggests three possible scenarios \cite{Swift2}. One is that $v_c$ depends only on the frequence via Eq.(\ref{e.vc}), as appears in our experiments. Second option is that it depends only on the geometrical dimension $D$. This seems to be the case for dc flows as summarized, {\it e.g.}, in Refs.\cite{Swanson,Donnelly}. The last option is that $v_c$ could depend on both $\omega$ and $D$. In this case the square-root dependence would show up only in a particular experiment where the frequency and dimensions are both scaled properly.  

While the frequency dependence of $v_c$ can be found from scaling no information on the absolute values or the temperature dependence can be inferred. As pointed out earlier \cite{HS} the prefactor 2.85 of our experimental result on oscillating spheres $v_c=2.85\,\sqrt{\kappa\omega}$ (see Fig.\ref{f:vcsphere}) may have some geometry dependence because the superfluid velocity field in general varies over the surface of the oscillating body. Hence, this prefactor can only be determined from experiment. In fact, the vibrating wire data of the Osaka group give a slightly smaller value 
\cite{Yano}. On the other hand, $v_c$ from the second-sound data of Kotsubo and Swift is about twice of what we find from our spheres at the same frequency and temperature. 

Another difference  concerns the temperature dependence of $v_c$. Kotsubo and Swift find that their $v_c$ decreases with rising temperature from 1.2 K to 2.0 K, while in our case we observe a 10\% increase from 1.0 K to 1.9 K that follows from the Vinen equation, see below. Possibly, these differences may be due to the very different mechanisms by which the oscillating flows are excited. 

%\bigskip

\section{Conclusions}

In our earlier work \cite{arxiv,HS,HS2} we have presented two different ways of deriving the frequency dependence of $v_c$. A qualitative but very general argument was based on the superfluid Reynolds number $Re_s = l v/\kappa \sim 1$ by identifying the oscillation amplitude $v/\omega$ as the characteristic length scale $l$, which immediately implies the square-root behaviour of $v_c(\omega)$.

Secondly, and in more detail, we started from the Vinen equation \cite{VinenEq}, which was established for dc flow and recently modified by Kopnin \cite{Kolya}

\begin{equation}
dL/dt = b\,(v L^{3/2} - \kappa L^2),\\
\label{e.vinen}
\end{equation}
where $b = A(1-\alpha')-B\alpha$ and $A,B \sim 1 $. (Equation (\ref{e.vinen}) does not contain the logarithmic term $\beta$ and is invariant under the same scaling as given in the Introduction.) Applying the Vinen equation to ac flow we have postulated that the vortex tangle, in order to follow the amplitude $v$ of the ac flow field, must have a relaxation time $\tau = 2\kappa / b v^2$ \cite{Kolya} that is shorter than the period of the velocity field, i.e., $\omega \, \tau < 1$. This gives $v_c(\omega) \sim \sqrt{2\kappa\omega/b} $. Below 1 K mutual friction can be neglected, hence $b \sim 1 $. Towards higher temperatures $b$ gradually decreases and hence $v_c$ increases. This is in agreement with our data \cite{HS2}. Clearly, both methods have a qualitative character. But to our knowledge, a theory of oscillatory superflow, in particular an extension of the Vinen equation to ac flow is not yet available.

In contrast, our present approach is based on dynamical scaling that follows from symmetry properties of the equations of vortex dynamics. Equation (\ref{e.vc}) is a rigorous result. The logarithmic terms contain the scaling of the length which in our experiments, however, can be neglected. Therefore, the square-root dependence of $v_c(\omega)$ is well established.  

\begin{acknowledgements}
RH was supported by the Academy of Finland (grant 218211).
\end{acknowledgements}


\begin{thebibliography}{}
% etc

\bibitem{Frisch} U. Frisch, \textsl{Turbulence}
(Cambridge University Press, Cambridge, 1995).

\bibitem{Schwarz}
K.W. Schwarz, Phys. Rev. B \textbf{38}, 2398 (1988). 

%\bibitem{Rozen}
%K.W. Schwarz and J.R. Rozen, Phys. Rev. B \textbf{44}, 7563 (1991), Eq.(11).

\bibitem{VinenEq}
W.F. Vinen, Proc. Roy. Soc. London, Ser. A \textbf{242}, 493 (1957).

\bibitem{Swift1}
V. Kotsubo and G.W. Swift, Phys. Rev. Lett. \textbf{62}, 2604 (1989).
 
\bibitem{Swift2}
V. Kotsubo and G.W. Swift, J. Low Temp. Phys. \textbf{78}, 351 (1990). 
 
\bibitem{Schwarz2}
K.W. Schwarz, Phys. Rev. B \textbf{18}, 245 (1978).

\bibitem{arxiv}
R. H\"anninen and W. Schoepe, arXiv:0801.2521 [cond-mat.other].

\bibitem{Swanson}
C.E. Swanson and R.J. Donnelly, J. Low Temp. Phys. \textbf{61}, 363 (1985). 

\bibitem{Donnelly} R.J. Donnelly, \textsl{Quantized Vortices in Helium II} 
(Cambridge University Press, Cambridge, 1991).

\bibitem{HS}
R. H\"anninen and W. Schoepe, J. Low Temp. Phys. \textbf{158}, 410 (2010). 

\bibitem{Yano}
H. Yano, private communication.
 
\bibitem{HS2}
R. H\"anninen and W. Schoepe, J. Low Temp. Phys. \textbf{153}, 189 (2008). 
 
\bibitem{Kolya}
N.B. Kopnin, Phys. Rev Lett. \textbf{92}, 135301 (2004). 

\end{thebibliography}


\end{document}